\documentclass[10pt]{article}

\usepackage{amsmath}
\usepackage{amssymb}
\usepackage{latexsym}
\usepackage{amscd}

\newcommand{\dif}{\mathrm{d}} 
\newcommand{\eps}{\varepsilon}
\newcommand{\QED}{\raisebox{0.5ex}{\framebox[1.5ex]{\rule{0ex}{0ex}}}}
\newcommand{\R}{\mathbf{R}} 
\newcommand{\vol}{\mathrm{Vol}_{g}}
\newcommand{\mychi}{\raisebox{0.5ex}{$\chi$}}
\renewcommand{\div}{\mathrm{div}} 
\newcommand{\Ric}{\mathit{Ric}}
\newcommand{\tjacobi}{traceless Jacobi } 
\newcommand{\jacobi}{Jacobi }

\allowdisplaybreaks

\begin{document}

\title{Exploring the Conformal Constraint Equations}

\author{Adrian Butscher \\ 
Max Planck Institute for Gravitational Physics \\ E-mail: \ttfamily 
butscher@aei.mpg.de}

\maketitle

\renewcommand{\baselinestretch}{0.9}
\normalsize

\begin{abstract}
    One method of studying the asymptotic structure of spacetime is to
    apply Penrose's conformal rescaling technique.  In this setting,
    the rescaled Einstein equations for the metric and the conformal
    factor in the unphysical spacetime degenerate where the conformal
    factor vanishes, namely at the boundary representing null
    infinity.  This problem can be avoided by means of a technique of
    H. Friedrich, which replaces the Einstein equations in the
    unphysical spacetime by an equivalent system of equations which is
    regular at the boundary.  The initial value problem for these
    equations produces a system of constraint equations known as the
    conformal constraint equations.  This work describes some of the
    properties of the conformal constraint equations and develops a
    perturbative method of generating solutions near Euclidean space
    under certain simplifying assumptions.
\end{abstract}

\renewcommand{\baselinestretch}{1.1}
\normalsize

\section{Introduction}

A model for the asymptotic structure of spacetime was suggested by
Roger Penrose in \cite{penrose1} (see also \cite{jorg} for a review of
the development of these ideas) using the technique of conformal
rescaling.  Since the reader is by now familiar with the details of
the conformal rescaling construction, only enough will be said here to
fix the notation to be used in the remainder of this article.  The
object under study will consist of an \emph{asymptotically simple
spacetime}: that is, a \emph{physical spacetime} consisting of a
smooth, time- and space-orientable Lorentz manifold $\tilde{M}$ with
metric $\tilde{g}$ satisfying the conditions:

\begin{enumerate}
    \item $\tilde{M}$ is diffeomorphic to the interior of an
    \emph{unphysical spacetime} $M$, which is a smooth Lorentz
    manifold $M$ with metric $g$ that has a boundary $\partial M$;
    
    \item there is a smooth function $\Omega : M \rightarrow \R$, the
    \emph{conformal factor}, which satisfies $\tilde{g} = \Omega^{-2}
    g$ on $\tilde{M}$ (the pull-back by the diffeomorphism $\tilde{M}
    \rightarrow int(M)$ has been suppressed for convenience);
    
    \item $\Omega = 0$ but $\dif \Omega \neq 0$ on $\partial M$;
    
    \item every null geodesic on $\tilde{M}$ acquires a future and 
    past endpoint on $\partial M$.
    
\end{enumerate}
Furthermore, $(\tilde{M}, \tilde{g})$ will be assumed to satisfy
Einstein's vacuum equation $\mathit{Ric}(\tilde{g}) = 0$.

The conformal boundary $\partial M$ describes \emph{null infinity} by
virtue of condition (4), and asymptotic properties of the physical
spacetime in null directions can be examined by studying the
properties of the unphysical spacetime near its boundary.  To this
end, one could use the fact that, due to the conformal equivalence
with the physical spacetime, the quantities $\Omega$ and $g$ must
satisfy the conformally rescaled version of Einstein's equation,
namely that $\mathit{Ric} (\Omega^{-2} g) = 0$.  However, this
equation has the drawback that it is degenerate near the boundary of
$M$ because there $\Omega \rightarrow 0$, and is thus not ideally
suited for analytic investigations of the nature of the spacetime at
null infinity.  One possible means of avoiding this difficulty is to
use a technique developed by Friedrich \cite{fried1}, which aims to
describe the geometry of the unphysical spacetime by means of a new,
yet fully equivalent system of equations derived from the equation
$\mathit{Ric} (\Omega^{-2}g) = 0$ that is formally regular at the
boundary of the unphysical spacetime.  These equations involve $g$,
$\Omega$ and several additional quantities and are known as the
\emph{conformal Einstein equations}.

As with Einstein's equations in the physical spacetime, it is possible
to attempt to solve the conformal Einstein equations in the unphysical
spacetime by means of an initial value formulation, where appropriate
initial data are defined on a spacelike hypersurface $\mathcal{Z}$ in
$M$ and then evolved in time.  Again as in the physical spacetime, the
conformal equations induce certain constraint equations on the initial
data; these equations are known as the \emph{conformal constraint
equations} and consist of a complicated system of coupled nonlinear
differential equations for the induced metric $h$ and second
fundamental form $\mychi$ of $\mathcal{Z}$, the conformal factor
restricted to $\mathcal{Z}$, and several additional quantities.  A
particular case of interest is when $\mathcal{Z}$ is asymptotically
\emph{hyperboloidal}, i.~e.\ $\mathcal{Z}$ intersects $\partial M$
transversely.  In this case, the evolution of the boundary of
$\mathcal{Z}$ forward in time produces the conformal boundary of the
unphysical spacetime, and global questions concerning the existence of
classes of spacetimes satisfying the definition of asymptotic
simplicity can be addressed.  See \cite{jorg} or \cite{fried3,fried4}
for a review of these ideas.

The purpose of this article is twofold.  First, it is to introduce the
conformal constraint equations and to investigate some of their
properties, which will be done in Section 2.  It will be found that,
in a certain sense, they describe in a coupled way two mathematical
problems --- namely, the elliptic boundary value problem for the
conformal factor $\Omega$ and the constraint problem arising from the
Gauss-Codazzi equations of $\mathcal{Z}$.  Furthermore, a simple
geometric assumption will be shown to lead to a special case of the
equations in which the first problem does not appear and the second is
in the forefront.  In this special case, the full system of conformal
constraint equations reduces to a much simpler and smaller system of
equations that will be called the \emph{extended constraint equations}
because they will turn out to be equivalent to the usual vacuum
Einstein constraint equations satisfied by the metric and second
fundamental form of $\mathcal{Z}$.  (Tackling the boundary value
problem is at present beyond the scope of this article but will be
considered in the future.)

The second goal of this article is to set up a perturbative approach
for generating solutions of the extended constraint equations in the
neighbourhood of a known solution, but only in the case of
time-symmetric data --- the more general case will be handled in
another future paper \cite{me4}.  This task will be accomplished in
Section 3 and the main theorem proved in this section appears on page
\ref{page:mainthm}.  Because the extended constraint equations are
equivalent to the usual constraint equations, the Main Theorem can be
interpreted as a new way of finding solutions of these equations, and
furthermore, it will turn out to be one that is completely different
from the `classical' (i.~e.\ Lichnerowicz-York) method of handling
them.  This issue will be discussed further in the Section 3.

\section{The Conformal Constraint Equations}

\subsection{Deriving the Equations}

\newcommand{\DBox}{\mbox{\raisebox{-0.25ex}{\large$\Box$}}}

Suppose $(M,g,\Omega)$ is an unphysical spacetime satisfying the
assumptions of asymptotic simplicity and thus that the metric and
conformal factor satisfy the rescaled version of Einstein's equation
\begin{equation}
    \label{eqn:moreblah}
    \mathit{Ric}(\Omega^{-2} g) = 0 \, .
\end{equation}
This section sketches briefly how equation \eqref{eqn:moreblah} for
$g$ and $\Omega$ leads first to the conformal Einstein equations for
$g$, $\Omega$ and additional quantities, and then to the conformal
constraint equations.  Begin by expanding \eqref{eqn:moreblah} to
obtain
\begin{equation}
    R_{\mu \nu} = -\frac{\DBox \Omega }{\Omega} \, g_{\mu \nu} -
    \frac{2}{\Omega} \, \nabla_{\mu} \nabla_{\nu} \Omega + \frac{3 \,
    \nabla^{\lambda} \Omega \nabla_{\lambda} \Omega}{\Omega^{2}} \,
    g_{\mu \nu} \, ,
    \label{eqn:confric}
\end{equation}
where $R_{\mu \nu}$ are the components of the Ricci tensor in the
unphysical spacetime, $\nabla_{\mu}$ is the covariant derivative of
the four-metric and $\DBox$ is its D'Alembertian operator.  Notice
that, as it is written, equation \eqref{eqn:confric} contains terms
with negative powers of $\Omega$ which tend to infinity near the
boundary of the unphysical spacetime.  Alternatively, if the equation
is multiplied through by $\Omega^{2}$, then the principal parts of the
differential operators acting on $g$ and $\Omega$, would tend to zero
at the boundary.  Either way, equation \eqref{eqn:confric} degenerates
near the boundary of the unphysical spacetime, making it an unwieldy
choice for studying the geometry of the spacetime near null infinity.

Helmut Friedrich has developed a procedure for obtaining a system of
equations equivalent to the rescaled Einstein equations
\eqref{eqn:confric} but that is formally regular at the boundary of
the unphysical spacetime and thus avoids the problems outlined above. 
This work can be found in several papers, see for example
\cite{fried1}.  Friedrich's derivation proceeded in the following way. 
Let $C_{\mu \nu \lambda \rho}$ be the Weyl tensor of the metric $g$
and define the quantities
\begin{equation}
    \begin{aligned}
	L_{\mu \nu } &= \frac{1}{2} R_{\mu \nu} - \frac{1}{12} R g_{\mu
	\nu} \\
	S_{\mu \nu \lambda \rho} &= \Omega^{-1} C_{\mu \nu \lambda
	\rho} \\
	\psi &= \frac{1}{4} \DBox \Omega + \frac{1}{24} R \Omega \, .
    \end{aligned}
    \label{eqn:defns}
\end{equation}
The tensor $S_{\mu \nu \lambda \rho}$ is smooth on $\partial M$
because under the assumptions of asymptotic simplicity, Penrose has
shown that $C_{\mu \nu \lambda \rho}$ vanishes at the boundary of $M$
\cite{penrose2} (a further condition on the topology of $\partial M$
--- that it admit spherical sections --- is also needed, and will be
assumed to hold).  Friedrich then found that the system of equations
\begin{equation}
    \label{eqn:confeinstein}
    \begin{gathered}
	\nabla_{\mu} \nabla_{\nu} \Omega = - \Omega L_{\mu \nu} + \psi
	g_{\mu \nu} \\
	\nabla_{\mu} \psi = - L_{\mu \nu} \nabla^{\nu} \Omega \\
	\nabla_{\lambda} L_{\mu \nu} - \nabla_{\mu} L_{\lambda \nu} =
	\nabla^{\rho} \Omega \, S_{\mu \lambda \nu \rho} \\
	\nabla^{\rho} S_{\mu \lambda \nu \rho} = 0 \\
	2 \Omega \psi - \nabla_{\mu} \Omega \nabla^{\mu} \Omega = 0 \\
	R_{\mu \nu \lambda \rho} = \Omega S_{\mu \nu \lambda \rho} +
	g_{\mu [ \lambda} L_{\nu ] \rho} - L_{\mu [ \lambda } g_{\nu ]
	\rho} \, ,
    \end{gathered}
\end{equation} 
where $R_{\mu \nu \lambda \rho}$ is the Riemann curvature tensor of
the metric $g$, can be derived from \eqref{eqn:confric}.  This is done
by rephrasing \eqref{eqn:confric} in terms of the quantities
\eqref{eqn:defns}, adjoining the Bianchi identity for the physical
spacetime and the unphysical spacetime, and adjoining the well-known
decomposition of the curvature tensor given by
$$R_{\mu \nu \lambda \rho} = C_{\mu \nu \lambda \rho} + g_{\mu
\lambda} L_{\nu \rho} - g_{\mu \rho} L_{\nu \lambda} + g_{\nu \rho}
L_{\mu \lambda} - g_{\nu \lambda} L_{\mu
\rho}\, .$$
Then the algebraic properties of the equations and the defined
quantities allows them to be manipulated into the regular ones listed
in \eqref{eqn:confeinstein}, which are known as the \emph{conformal
Einstein equations}.  The equivalence of \eqref{eqn:confeinstein} to
\eqref{eqn:confric} is confirmed as follows.  Suppose the quantities
$L$, $S$ and $\psi$ as well as $g$ and $\Omega$ satisfy
\eqref{eqn:confeinstein}.  Then by algebra, it can be shown that the
pair $(g,\Omega)$ satisfies \eqref{eqn:confric} and that $L$, $S$ and
$\psi$ relate to $\Omega$ and the curvature quantities in the manner
indicated in \eqref{eqn:defns}.  (The algebra is fairly
straightforward: for instance, the last equation in
\eqref{eqn:confeinstein} identifies $L$ and $S$ as components of the
curvature tensor; then it is a matter of computation to recover
equation \eqref{eqn:confric} from the remaining five.)

It is immediately clear that the equations in \eqref{eqn:confeinstein}
are regular when $\Omega = 0$.  Furthermore, not only do the conformal
Einstein equations contain the rescaled vacuum Einstein equations, but
they also contain the Bianchi identity for the curvature tensor,
though expressed in the new unknowns.  Thus one can consider
\eqref{eqn:confeinstein} to contain \emph{compatibility conditions}
since the Bianchi identity is in some sense a compatibility condition
for the curvature tensor --- meaning that the Bianchi identity is a
result of requiring second covariant derivatives to commute properly
(this can best be seen explicitly by rewriting the curvature tensor in
terms of the vector-valued connection 1-forms as in \cite{boothby},
whereby the Bianchi identity becomes an incarnation of the identity
$\dif^{2} = 0$ satisfied by the exterior differential operator).

Suppose now that $\mathcal{Z}$ is a spacelike hypersurface in $M$. 
The fact that the conformal Einstein equations constrain certain
initial data on $\mathcal{Z}$ can be seen by performing a $3+1$
splitting of spacetime near $\mathcal{Z}$.  Choose a frame $E_{a}$,
$a= 1,2,3$, for the tangent space of $\mathcal{Z}$ and complete this
to a frame for the unphysical spacetime by adjoining the
forward-pointing unit normal vector field $n$ of $\mathcal{Z}$.  Use
this frame to decompose the equations \eqref{eqn:confeinstein} into
components parallel and perpendicular to $\mathcal{Z}$.  The
constraint equations induced by the conformal Einstein equations are
those equations in which no second normal derivatives of $g$ or
$\Omega$, and no first normal derivatives of $L$, $S$ or $\psi$
appear.  The initial data are the unknown quantities which are found
in these equations; they are:
\begin{itemize}
    \item the induced metric of $\mathcal{Z}$, which will still be
    called $g$ (no confusion will arise because the 4-dimensional
    setting will not be considered further in the remainder of the
    this article),
    \item the second fundamental form $\mychi$ of $\mathcal{Z}$,
    \item the function $\Omega$ restricted to $\mathcal{Z}$,
    \item the normal derivative $n(\Omega) \big|_{\mathcal{Z}}$, to 
    be denoted $\Sigma$,
    \item the tensors $L_{ab} = E^{\mu}_{a} E^{\nu}_{b} L_{\mu \nu}$
    and $L_{a} = n^{\mu} E_{a}^{\nu} L_{\mu \nu}$,
    \item the tensors $\bar{S}_{abc} = n^{\mu} E^{\nu}_{a}
    E^{\lambda}_{b} E^{\rho}_{c} S_{\nu \mu \lambda \rho}$ and $S_{ab}
    = n^{\mu} n^{\nu} E_{a}^{\lambda} E_{b}^{\rho} S_{\lambda \mu \rho
    \nu}$,
    \item and the function $\psi$ restricted to $\mathcal{Z}$.
\end{itemize}
The constraint equations arising from the $3+1$ splitting are:
\begin{equation}
    \label{eqn:confcon}
    \begin{gathered}
	\nabla_{a} \nabla_{b} \Omega = \Sigma \mychi_{ab} - \Omega
	L_{ab} + \psi g_{ab} \\
	\nabla_{a} \Sigma = \mychi^{c}_{a} \nabla_{c} \Omega - \Omega
	L_{a} \\
	\nabla_{a} \psi = - \nabla^{b} \Omega L_{ba} - \Sigma L_{a} \\
	\nabla_{a} L_{bc} - \nabla_{b} L_{ac} = \nabla^{e}\Omega
	S_{ecab} - \Sigma S_{cab} - (\mychi_{ac} L_{b} - \mychi_{bc}
	L_{a}) \\
	\nabla_{a} L_{b} - \nabla_{b} L_{a} = \nabla^{e} \Omega \,
	S_{eab} + \mychi^{c}_{a} L_{bc} - \mychi^{c}_{b} L_{ac} \\
	\nabla^{a} \bar{S}_{abc} = \mychi^{a}_{b} S_{ac} -
	\mychi^{a}_{c} S_{ab} \\
	\nabla^{a} S_{ab} = \mychi^{ac} \bar{S}_{abc} \\
	0 = 2 \Omega \psi + \Sigma^{2} - \Vert \nabla \Omega \Vert^{2}
	\\
	\nabla_{c} \mychi_{ba} - \nabla_{b} \mychi_{ca} = \Omega
	\bar{S}_{abc} + g_{ab} L_{c} - g_{ac} L_{b} \\
	R_{ab} = \Omega S_{ab} + L_{ab} + \frac{1}{4} L^{c}_{c} g_{ab}
	- \mychi^{c}_{c} \mychi_{ab} + \mychi_{ca} \mychi^{c}_{b}
    \end{gathered}
\end{equation}
where $\nabla$ now denotes the covariant derivative operator on
$\mathcal{Z}$ corresponding to its induced metric $g$ and $R_{ab}$ is
its Ricci curvature.  These equations are known as the \emph{conformal
constraint equations}.  The derivation of these equations will not be
reproduced here --- the reader is asked to consult \cite{fried1} for
this material.  However, it is fairly easy to recognize the origin of
the various terms appearing above.  For example, the first two
equations arise as the tangential and tangential-normal components of
the first equation of \eqref{eqn:confeinstein}.  Furthermore, and more
importantly for the sequel, the last two equations arise as the Gauss
and Codazzi equations applied to the decomposition of the curvature
tensor given by the last equation of \eqref{eqn:confeinstein}.

The various tensor quantities that appear in \eqref{eqn:confcon}
possess certain symmetries as a result of their origin as components
of the curvature tensor: $L_{ab}$ is symmetric; $S_{ab}$ is symmetric
and trace-free; and $\bar{S}_{abc}$ is antisymmetric on its last two
indices, satisfies the Jacobi symmetry $\bar{S}_{abc} + \bar{S}_{cab}
+ \bar{S}_{bca} = 0$ and is trace-free on all its indices.  (Tensors
with these symmetries will appear often in the sequel.  Tensors of
rank three that are antisymmetric on their last two indices and
satisfy the Jacobi symmetry will be called \jacobi tensors for short
while those which are in addition trace-free will be called \tjacobi
tensors.)  Note that even though the tensor $S_{abcd} = E_{a}^{\mu}
E_{b}^{\nu} E_{c}^{\lambda} E_{d}^{\rho} S_{\mu \nu \lambda \rho}$
appears in the constraint equations, it is not a truly independent
initial datum because, thanks to the symmetries of $S_{\mu \nu \lambda
\rho}$, it can be written as $S_{abcd} = g_{a[c}S_{d]b} - S_{a[c}
g_{d]b}$.

The system \eqref{eqn:confcon} is clearly exceedingly complicated
because it is quasi-linear and highly coupled.  However, the advantage
provided by \eqref{eqn:confcon} is once again that it is formally
regular at the boundary of $\mathcal{Z}$.  For the sake of comparison,
recall the interior of $\mathcal{Z}$ can be viewed as a spacelike
hypersurface of the physical spacetime, and as such, satisfies the
usual Einstein constraint equations there.  In other words, if its
induced metric is denoted by $\tilde{g}$ and its second fundamental
form by $\tilde{\mychi}$, then
\begin{equation}
    \begin{gathered}
	\tilde{\nabla}^{a} \tilde{\mychi}_{ab} - \tilde{\nabla}_{b}
	\tilde{\mychi}^{a}_{a} = 0 \\
	\tilde{R} + (\tilde{\mychi}^{a}_{a})^{2} - \tilde{\mychi}^{ab}
	\tilde{\mychi}_{ab} = 0 \, , 
    \end{gathered}
    \label{eqn:usualcon}
\end{equation}
where $\tilde{\nabla}$ is the covariant derivative operator of the
metric $\tilde{g}$ and $\tilde{R}$ is its scalar curvature.  These
equations can be rephrased in terms of $g$, $\mychi$ and $\Omega$ in
the unphysical spacetime by conformal transformation.  The necessary
transformation rules are that $\tilde{g} = \Omega^{-2} g$ and also
that $\tilde{\mychi} = \Omega^{-1} \mychi + \Sigma \Omega^{-2} g$
(which can be found by conformally transforming the definition of the
second fundamental form as the normal component of the covariant
derivative restricted to $\tilde{\mathcal{Z}}$).  The resulting
equations are
\begin{equation}
    \label{eqn:confusualcon}
    \begin{gathered}
	\Omega^{2} \big( R + (\mychi^{a}_{a})^{2} - \mychi^{ab}
	\mychi_{ab} \big) + 4 \Omega \Delta_{g} \Omega - 6 \Vert
	\nabla \Omega \Vert^{2} + 4 \Omega \Sigma \mychi^{a}_{a} +
	6\Sigma ^{2} =0 \\
	\Omega \big( \nabla_{a} \mychi^{a}_{b} - \nabla_{b}
	\mychi^{a}_{a} \big) - 2 \nabla_{b} \Sigma - 2 \mychi^{a}_{b}
	\, \
nabla_{a} \Omega = 0\, ,
    \end{gathered}
\end{equation}
where $\Sigma = n(\Omega) \big|_{\mathcal{Z}}$ and $\Delta_{g}$ is the
Laplacian of the metric $g$.  Once again, the principal parts of these
equations contain factors of $\Omega$ and thus degenerate as $\Omega
\rightarrow 0$ near the boundary of $\mathcal{Z}$.  This behaviour
does not arise in the conformal constraint equations.

The conformal constraint equations listed in \eqref{eqn:confcon} are
equivalent to the usual constraint equations \eqref{eqn:confusualcon}
because if $(g,\mychi,\Omega, \Sigma)$ solves \eqref{eqn:confusualcon}
and the subsidiary quantities $S$, $\bar{S}$, $L$ and $\psi$ are
defined as indicated in \eqref{eqn:confcon} (e.~g.\ the last equation
defines $\psi$; then the first equation defines the 2-tensor $L_{ab}$,
etc.\ ), then straightforward computation shows that the conformal
constraint equations are satisfied; furthermore, if $(g, \mychi,
\Omega, \Sigma, S, \bar{S}, L, \psi)$ satisfies \eqref{eqn:confcon},
then it can be shown that $(g, \mychi, \Omega, \Sigma)$ satisfies
\eqref{eqn:confusualcon}, and consequently, $\tilde{g}$ and
$\tilde{\mychi}$, given by the transformation rules above, satisfy the
usual constraint equations \eqref{eqn:usualcon}.  These considerations
thus suggest one method for constructing solutions of the conformal
constraint equations: construct any solution $(\tilde{g},
\tilde{\mychi})$ of the usual constraint equations using standard
techniques, choose a conformal factor, perform the transformations to
the unphysical spacetime and use the conformal constraint equations to
define the subsidiary quantities in terms of $(\tilde{g},
\tilde{\mychi})$.  Then these new quantities satisfy the conformal
constraint equations.

Consequently, it is possible to assume the existence of initial data
with well-defined asymptotic properties (essentially given by the
transformation rules above) and study only the time evolution of the
data according to the conformal Einstein equations
\eqref{eqn:confeinstein}.  This is the idea behind the work of
Friedrich in \cite{fried1} (extended in \cite{fried5}), where the time
evolution of suitably small initial data on an asymptotically
hyperboloidal hypersurface was studied and a complete future
development was found.  The nature of the asymptotic structure of this
class of solutions near null infinity, and in particular the
relationship between the asymptotic structure of the solution and the
asymptotic structure of the initial data, was then analyzed
extensively by Andersson, Chru\'sciel and Friedrich in \cite{acf}
(extended by Andersson and Chru\'sciel in \cite{ac1,ac2}), and was
based on the rescaled Einstein equations \eqref{eqn:confeinstein} and
their constraints \eqref{eqn:confusualcon}.  However, the problem of
the vanishing of the conformal factor near the boundary of the
unphysical spacetime and the resultant degeneration of these equations
remains a part of the ACF methods.  Thus they are not ideally suited
for certain applications, in particular for implementing numerical
studies of asymptotically hyperboloidal data near null infinity where
the presence of negative powers of $\Omega$ can cause computational
codes to crash (see \cite{jorg} for details).  It is for this reason
that new methods for solving \eqref{eqn:confcon} directly, rather than
through the usual constraint equations, must be developed.  This
question will begin to be tackled in the remainder of this article.

\subsection{Reduction to the Extended Constraint Equations}

The complexity of the conformal constraint equations makes it a
daunting task to attempt to develop any methods for obtaining
solutions of the equations in their full generality.  However, a great
deal of structure is contained within these equations, and the hope is
that this structure can be exploited in the search for solutions.  For
instance, it is possible to disentangle in some sense the equations
relating to the conformal factor and its associated boundary value
problem from the equations related to the Gauss-Codazzi equations of
$\mathcal{Z}$ by restricting to a special case of the equations.

The special case that will be considered in the rest of this article
is to assume that the conformal diffeomorphism between $\tilde{M}$ and
$M$ is the identity, and consequently that the conformal factor is
trivial (i.~e.\ $\Omega=1$) in the unphysical spacetime.  This is
somewhat of a strange simplification, because it requires that the
spacetime $M$ have empty boundary (since $\Omega^{-1}(0) = \partial
M$)!  One would thus not find oneself in this special case in practice
since the whole point of the conformal constraint equations is to
study hyperboloidal initial data in a conformally rescaled spacetime
that has a boundary at null infinity.  Nevertheless, the
simplification afforded by the assumption $\Omega = 1$ is worthwhile
to consider from a mathematical point of view because it will
accomplish the disentanglement described above and allow the
Gauss-Codazzi-type equations within the conformal constraint equations
to be studied in isolation.

To see this explicitly, one must substitute $\Omega = 1$ and $\Sigma =
0$ (which is consistent with the assumption that $\Omega = 1$ in
spacetime since $\Sigma = n(\Omega) \big|_{\mathcal{Z}} = 0$ where $n$
is the forward-pointing unit normal of $\mathcal{Z}$) into the
equations \eqref{eqn:confcon}.  One first sees that $L_{ab}$, $L_{a}$
and $\psi$ are forced to vanish under this assumption, and then that
the conformal constraint equations reduce to the following system of
four coupled equations:
\begin{equation*}
\begin{gathered}
    R_{ab} = S_{ab} - \mychi^{c}_{c} \mychi_{ab} + \mychi^{c}_{a}
    \mychi_{cb} \\
    \nabla_{c} \mychi_{ab} - \nabla_{b} \mychi_{ac} = \bar{S}_{abc} \\
    \nabla^{a} \bar{S}_{abc} = \mychi^{a}_{b} S_{ac} - \mychi^{a}_{c}
    S_{ab} \\
    \nabla^{a} S_{ab} = - \mychi^{ac} \bar{S}_{abc} \, .
\end{gathered}
\end{equation*}
Here, covariant derivatives are taken with respect to the induced
metric $g_{ab}$ of $\mathcal{Z}$ and $\mychi_{ab}$ is the second
fundamental form of $\mathcal{Z}$.  As before, the tensor $S_{ab}$ is
symmetric and trace-free with respect to $g_{ab}$ whereas the tensor
$\bar{S}_{abc}$ is a traceless Jacobi tensor.  These four quantities
are the unknowns for which these equations must be solved.  For
reasons that will become apparent later on, it will be helpful work
instead with the equivalent system obtained by replacing $S_{ab}$ and
$S_{ac}$ in the third equation by $R_{ab}$ and $R_{ac}$ from the first
equation.  The system one obtains is actually just
\begin{equation} \label{eqn:cces}
\begin{gathered}
    R_{ab} = S_{ab} - \mychi^{c}_{c} \mychi_{ab} + \mychi^{c}_{a}
    \mychi_{cb} \\
    \nabla_{c} \mychi_{ab} - \nabla_{b} \mychi_{ac} = \bar{S}_{abc} \\
    \nabla^{a} \bar{S}_{abc} = \mychi^{a}_{b} R_{ac} - \mychi^{a}_{c}
    R_{ab} \\
    \nabla^{a} S_{ab} = - \mychi^{ac} \bar{S}_{abc} \, ,
\end{gathered}
\end{equation}
because the terms cubic in $\mychi$ vanish.

Notice that because of the symmetries of $S$ and $\bar{S}$, if the
traces of the first two equations of \eqref{eqn:cces} are taken, then
the usual constraint equations \eqref{eqn:usualcon} result. 
Furthermore, if $g_{ab}$ and $\chi_{ab}$ satisfy the usual constraint
equations and one \emph{defines} $\bar{S}_{abc}$ and $S_{ab}$ by the
first two equations of \eqref{eqn:cces} respectively, then the
remaining two equations follow by straightforward algebra and the
Bianchi identity.  Thus equations \eqref{eqn:cces} are equivalent to
the usual vacuum Einstein constraint equations and for this reason are
called the \emph{extended constraint equations}.

\subsection{Properties of the Extended Constraint Equations}

The extended constraint equations \eqref{eqn:cces} are clearly
formally much simpler than the full system of conformal constraint
equations.  However, several essential features of the full equations
remain.  These features refer to the ellipticity properties of the
various differential operators appearing in \eqref{eqn:cces} as well
as to the compatibility conditions built into these equations.

\medskip \noindent \scshape Ellipticity Properties \upshape \medskip 

One must consider the principal symbols of the operators that appear
on the left hand sides of the extended constraint equations in order
to understand their ellipticity properties.  Begin with a definition
of the symbol.  Recall that if $P: C^{\infty} (\R^{n}, \R^{N})
\rightarrow C^{\infty}(\R^{n}, \R^{M})$ is a linear differential
operator of order $m$ with constant coefficients, then it can be
expressed as
$$P(u) = \sum_{\alpha_{1} + \cdots + \alpha_{n} = m} \left(
\sum_{i=1}^{N} b^{\alpha_{1} \cdots \alpha_{n}}_{i} \frac{\partial^{m}
u^{i} }{(\partial x^{1})^{\alpha_{1}} \cdots (\partial x^{n})
^{\alpha_{n}}} \right) + P_{0}(u) \, ,$$
where $P_{0}$ is a differential operator of order less than or equal
to $m-1$ and the $b^{\alpha_{1} \cdots \alpha_{n}}_{i}$ are elements
of $\R^{M}$.  The \emph{principal symbol} of $P$ is the family of
linear maps given by
$$\sigma_{\xi}(v) = \sum_{\alpha_{1} + \cdots + \alpha_{n} = m} \left(
\sum_{i=1}^{N} b^{\alpha_{1} \cdots \alpha_{n}}_{i}
\xi_{1}^{\alpha_{1}} \cdots \xi_{n}^{\alpha_{n}} v^{i} \right) $$
for any non-zero $(\xi_{1}, \ldots, \xi_{n}) \in \R^{n}$ and $v \in
\R^{N}$.  Furthermore, the operator $P$ is called
\emph{underdetermined elliptic} if the symbol is surjective for each
non-zero $\xi$, \emph{overdetermined elliptic} if the symbol is
injective for each non-zero $\xi$ and simply \emph{elliptic} if the
symbol is bijective for each non-zero $\xi$.  An operator with
non-constant coefficients has a symbol at each point of the domain,
while for a nonlinear operator, it is the linearization which has a
symbol at each given $u \in C^{\infty}(\R^{n}, \R^{N})$.  Such
operators are overdetermined, underdetermined or elliptic if their
symbols possess these properties uniformly.

To understand the ellipticity properties of the conformal constraint
equations, begin with the equation for the metric $g_{ab}$.  It is
quasi-linear in $g$, with highest-order terms given by
$$g_{ab} \mapsto g^{cd} \left( \frac{\partial^{2} g_{ad}}{\partial
x^{b} \partial x^{c}} + \frac{\partial^{2} g_{bd}}{\partial x^{a}
\partial x^{c}} - \frac{1}{2} \frac{\partial^{2} g_{ab}}{\partial
x^{c} \partial x^{d}} - \frac{1}{2} \frac{\partial^{2}
g_{cd}}{\partial x^{a} \partial x^{b} } \right) \, .$$
The linearization of this expression at a given metric is neither
over- nor underdetermined elliptic, nor is it elliptic.  However, it
is well known that the Ricci curvature is degenerate as an operator on
metrics because it is invariant under changes of coordinates of the
metric, and that the Ricci curvature operator can be made formally
elliptic by making an appropriate choice of coordinate gauge.  The
standard choice is to require that the metric be expressed in
\emph{harmonic coordinates}, which are defined by the requirement that
the coordinate functions $x^{a}$ are harmonic functions, i.~e.\ that
$\Delta_{h} x^{a} = 0$ for each $a$.  (Since the metric itself depends
on the coordinate functions, the requirement that the coordinates be
harmonic is in fact a nonlinear condition.  Nevertheless, the
existence of such coordinates, defined outside sufficiently large
balls in $\R^{3}$ for any asymptotically flat metric, has been
guaranteed by Bartnik in \cite{bartnik}.)

To show that the Ricci operator is elliptic in harmonic coordinates,
first note that a straightforward calculation implies that the
harmonic coordinate condition $\Delta_{g} x^{a} = 0$ for all $a$ is
equivalent to the condition $g^{bc} \Gamma_{bc}^{a} = 0$ for all $a$
on the Christoffel symbols of $g$.  Now set $\Gamma^{a} = g^{bc}
\Gamma_{bc}^{a}$ (and also $\Gamma_{a} = g_{as} \Gamma^{s}$), and then
recall that the components of the Ricci tensor satisfy
\begin{equation}
    R_{ab} = R^{H}_{ab} + \frac{1}{2} (\Gamma_{a;b} + \Gamma_{b;a})
    \label{eqn:redric}
\end{equation}
where $R^{H}_{ab}$ are the components of the \emph{reduced} Ricci
operator defined by
\begin{equation}
    R_{ab}^{H} = - \frac{1}{2} g^{rs} g_{ab,rs} + q(\Gamma) \, .
    \label{defn:redric}
\end{equation}
In the expressions above, a comma denotes ordinary differentiation
with respect to the coordinates, a semicolon denotes covariant
differentiation (since $\Gamma^{a}$ is not a tensor, this is to be
taken formally; i.~e.\ $\Gamma_{a;b} = \Gamma_{a,b} - \Gamma_{s}
\Gamma^{s}_{ab}$), and $q(\Gamma)$ denotes a term that is quadratic in
the components $\Gamma^{a}$.  The reduced Ricci operator is clearly
elliptic in $g$.  Since $\Gamma^{a} = 0$ for all $a$ in harmonic
coordinates, $R_{ab}(g) = R^{H}_{ab}(g)$ in these coordinates, and
thus the Ricci operator is elliptic in $g$ when $g$ satisfies the
harmonic coordinate condition.

The second equation in the extended constraint equations is linear in
$\mychi_{ab}$ and its left hand side defines a differential operator
$\mychi_{ab} \mapsto \nabla_{c} \mychi_{ab} - \nabla_{b} \mychi_{ac}$
from the space of symmetric tensors to the space of Jacobi tensors. 
(It can be easily verified that the left hand side of the first
equation in \eqref{eqn:cces} satisfies the relevant symmetries. 
However, it can also be verified that the left hand side is not
\emph{a priori} traceless on all its indices --- this is only a
requirement on the eventual solution since the left hand side is
equated with a traceless Jacobi tensor.)  The principal symbol of this
operator is
\begin{equation*}
    \sigma_{\xi} : \mychi_{ab} \mapsto \xi_{c} \mychi_{ab} - \xi_{b}
    \mychi_{ac} \, .
\end{equation*}
By the following simple argument, one can show that $\sigma_{\xi}$ has
a one-dimensional kernel and is not surjective.

Suppose first that $\sigma_{\xi} (\mychi_{ab}) = 0$ for some non-zero
$\xi$.  Since $\xi_{a} \xi^{a} \neq 0$, one can write uniquely
$\mychi_{ab} = \mychi_{ab}^{0} + c \xi_{a} \xi_{b}$ for some $c$,
where $\mychi_{ab}^{0}$ is trace-free.  Substituting this expression
for $\chi_{ab}$ yields
\begin{equation}
    \xi_{b} \mychi_{ac}^{0} - \xi_{c} \mychi_{ab}^{0} = 0 \, .
    \label{eqn:symbol}
\end{equation}
Taking the trace over $a$ and $b$ implies that $\xi^{c}
\mychi_{ac}^{0} = 0$.  Then, contracting with $\xi^{c}$ gives $\xi^{c}
\xi_{c} \mychi_{ab}^{0} = 0$, or $\mychi_{ab}^{0} = 0$.  Consequently,
the kernel of the symbol $\sigma_{\xi}$ is one-dimensional, and
consists of tensors of the form $c \xi_{a}\xi_{b}$.  Next, since the
space of symmetric 2-tensors is six-dimensional, the image of the
symbol is five-dimensional.  Now, the target space of \jacobi tensors
is eight-dimensional because any \jacobi tensor can be decomposed as
$T_{abc} = \eps^{e}_{\hspace{1ex} bc} F_{ae} + A_{b} g_{ac} - A_{c}
g_{ab}$ where $F_{ae}$ is a trace-free and symmetric tensor
(accounting for five dimensions), $A_{b}$ is a 1-form (accounting for
the remaining three), and $\eps_{abc}$ is the fully antisymmetric
permutation symbol.  The symbol can thus not be surjective.  Note,
however, that when it is restricted to trace-free tensors, the
principal symbol \emph{is} at least injective.  Consequently, the
first equation of \eqref{eqn:cces} is overdetermined elliptic when
restricted to the space of trace-free symmetric 2-tensors.

The third and fourth equations in \eqref{eqn:cces} are linear in
$\bar{S}_{abc}$ and $S_{ab}$ respectively.  It can be shown that the
operators $\bar{S}_{abc} \mapsto \nabla^{a} \bar{S}_{abc}$ and $S_{ab}
\mapsto \nabla^{a} S_{ab}$ are underdetermined elliptic by
demonstrating that their principal symbols $\bar{S}_{abc} \mapsto
\xi^{a} \bar{S}_{abc}$ and $S_{ab} \mapsto \xi^{a} S_{ab}$ are
surjective maps from the space of symmetric, trace-free tensors onto
the space of 1-forms and from the space of traceless Jacobi tensors
onto the space of antisymmetric 2-tensors, respectively.  These are
fairly straightforward calculations and left to the reader.

\medskip \noindent \scshape Compatibility Conditions \upshape \medskip

As mentioned in Section 2.1, the conformal Einstein equations
\eqref{eqn:confeinstein} contain the Bianchi identity, and was
interpreted as being a compatibility condition for the other
equations.  Such compatibility conditions are also to be found in the
conformal constraint equations; the present goal is to exhibit this
explicitly.  Begin by considering the first and fourth equations in
\eqref{eqn:cces}.  The Bianchi identity for the Ricci curvature is
$$\nabla^{a} R_{ab} - \frac{1}{2} \nabla_{b} R = 0 \, ,$$
whereby the first equation of \eqref{eqn:cces} implies
\begin{align}
    0 &= \nabla^{a} \big( S_{ab} - \mychi^{c}_{c} \mychi_{ab} +
    \mychi^{c}_{a} \mychi_{cb} \big) - \frac{1}{2} \nabla_{b} \big(- 
    (\mychi^{c}_{c})^{2} + \mychi^{ac} \mychi_{ac} \big) \notag \\
    &= \nabla^{a} S_{ab} - \big( \mychi^{c}_{c} \delta^{a}_{b} -
    \mychi^{a}_{b} \big) \big( \nabla^{u} \mychi_{au} - \nabla_{a}
    \mychi_{u}^{u} \big) - \mychi^{ca} \big( \nabla_{b} \mychi_{ac} -
    \nabla_{c} \mychi_{ab} \big) \notag \\
    &= \nabla^{a} S_{ab} - \big( \mychi^{c}_{c} \delta^{a}_{b} -
    \mychi^{a}_{b} \big) h^{uv} \bar{S}_{uav} + \mychi^{ac}
    \bar{S}_{abc}
    \label{eqn:vanish}
\end{align}
using the second equation in \eqref{eqn:cces} and its trace.  By the
symmetries of $\bar{S}_{abc}$, the middle term in \eqref{eqn:vanish}
vanishes, leaving
$$0 = \nabla^{a} S_{ab} + \mychi^{ac} \bar{S}_{abc} \, ,$$
which is exactly the fourth equation of \eqref{eqn:cces}.

The second and third equations of \eqref{eqn:cces} consist of a
constraint equation and its compatibility condition as well, but in a
different sense.  Recall that what a compatibility condition should
reflect is that second second covariant derivatives commute properly. 
Consider, then, the result of commuting the second covariant
derivatives of the second equation of \eqref{eqn:cces}.  Begin with
$$\bar{S}_{abc} = \nabla_{c} \mychi_{ab} - \nabla_{b} \mychi_{ac}$$
and compute
\begin{align}
    \eps^{ebc} \nabla_{e} \bar{S}_{abc} &= 2 \eps^{ebc} \nabla_{e}
    \nabla_{c} \mychi_{ab} \notag \\
    &= \eps^{ebc} \big( \nabla_{e} \nabla_{c} \mychi_{ab} - \nabla_{c}
    \nabla_{e} \mychi_{ab} \big) \notag \\
    &= \eps^{ebc} \big( R_{eca}^{\hspace{3ex} s} \mychi_{sb} +
    R_{ecb}^{\hspace{3ex} s} \mychi_{as} \big) \notag \\
    &= \eps^{ebc} R_{eca}^{\hspace{3ex} s} \mychi_{sb}
    \label{eqn:subme}
\end{align}
since the symmetries of $R_{abcd}$ imply that $\eps^{abc} R_{abcd} =
0$.  Now substitute in \eqref{eqn:subme} the well-known decomposition
of the curvature tensor in three dimensions, namely that
$$R_{eca}^{\hspace{3ex} s} = g_{ea} R^{s}_{c} - \delta^{s}_{e} R_{ca}
+ \delta^{s}_{c} R_{ea} - g_{ca} R^{s}_{e} - \frac{1}{2} R \big(
g_{ea} \delta^{s}_{c} - \delta^{s}_{e} g_{ca} \big) \, ,$$
to obtain
\begin{equation}
    \eps^{ebc} \nabla_{e} \bar{S}_{abc} = 2 \eps^{bc}_{\hspace{2ex} a}
    \mychi^{s}_{b} R_{cs} \, .
    \label{eqn:equiv}
\end{equation}

\noindent \scshape Claim: \upshape equation \eqref{eqn:equiv} is
exactly the third equation of \eqref{eqn:cces}.  To see this, recall
that a traceless Jacobi tensor can be decomposed as $\bar{S}_{abc} =
\eps^{e}_{\hspace{1ex} bc} F_{ae}$ where $F_{ae}$ is trace-free and
symmetric.  Consequently,
\begin{align}
    \eps^{ebc} \nabla_{e} \bar{S}_{abc} &= \eps^{ebc} \nabla_{e}
    \eps^{u}_{\hspace{1ex} bc} F_{au} \notag \\
    &= 2 \nabla^{e} F_{ae} \notag \\
    &= 2 \nabla^{e} F_{ea} \tag{by symmetry} \\
    &= \eps_{a}^{\hspace{1ex} bc} \nabla^{e} \eps^{u}_{\hspace{1ex} bc}
    F_{eu} \notag \\
    &= \eps_{a}^{\hspace{1ex} bc} \nabla^{e} \bar{S}_{ebc} \, .
    \label{eqn:symone}
\end{align}
Thus \eqref{eqn:symone} together with \eqref{eqn:equiv} implies that
$$\eps_{a}^{\hspace{1ex} bc} \nabla^{e} \bar{S}_{ebc} = 2 \eps^{bc}_{
\hspace{2ex} a} \mychi^{s}_{b} R_{cs} \, ,$$
which is the third equation of \eqref{eqn:cces} (or at least its dual,
but this is equivalent).

\section{Asymptotically Flat Solutions of the Extended Constraint
Equations in the Time Symmetric Case}

\subsection{Statement of the Main Theorem}

Because the conformal boundary of the spacetime $\tilde{M}$ is absent
under the triviality assumptions that have been made on the conformal
diffeomorphism, a natural setting in which to investigate the extended
constraint equations \eqref{eqn:cces} is the case in which $\tilde{M}$
is asymptotically Minkowski space and that $\mathcal{Z}$ is
asymptotically flat.  In fact, one solution of the extended constraint
equations satisfying these conditions is when $\mathcal{Z} = \R^{3}$
and the initial data is Euclidean metric $g = \delta$ with vanishing
second fundamental form and tensors $S$ and $\bar{S}$.  Neighbouring
asymptotically flat solutions are those whose metric $g$ is a small
perturbation of $\delta$ that decays suitably to $\delta$ near
infinity, and $\mychi$, $\bar{S}$ and $S$ are also small and decay
suitably.  These solutions are in addition \emph{time symmetric} if
their second fundamental form actually vanishes identically.  The
theorem that will be proved in the remainder of this article is a
characterization of the space of asymptotically flat \emph{and}
time-symmetric solutions of the extended constraint equations in the
neighbourhood of the trivial solution given above.  The case of
non-time-symmetric solutions is as yet beyond the scope of this
article, though a future paper by the Author will clear this up
\cite{me4}.

Under the assumption of time-symmetry, the requirement that $\mychi =
0$ implies that $\bar{S} = 0$ as well, and so the extended constraint
equations further reduce to the following system of equations:
\begin{equation*}
    \begin{gathered}
	\nabla^{a} S_{ab} = 0 \\
	R_{ab}(g) = S_{ab}  
    \end{gathered}
\end{equation*}
for the unknown metric $g$ and unknown trace-free and symmetric tensor
$S$.  Since these equations will be solved for metrics near the
Euclidean metric, it will be preferable to write metrics as small
perturbations of the Euclidean metric of the form $\delta + h$ where
$h$ is a symmetric tensor suitably near $0$.  Thus the above system
should be replaced with the system
\begin{equation}
    \label{eqn:simplextcon}
    \begin{gathered}
	\nabla^{a} S_{ab} = 0 \\
	R_{ab}(\delta + h) = S_{ab} \, .
    \end{gathered}
\end{equation}
The covariant derivative here corresponds to the metric $\delta + 
h$.  The theorem that will be proved is the following.

\label{page:mainthm}
\medskip \noindent \bfseries Main Theorem\mdseries: \itshape There
exists a Banach space $B$ of free data along with a neighbourhood $U$
of zero in $B$, Banach spaces $Y$ and $Y'$ of symmetric 2-tensors, and
smooth functions $\psi:U \rightarrow Y$ and $\psi':U \rightarrow Y'$
with $\psi(0) = \psi'(0) = 0$ so that for every $b \in U$, the
following hold:
\begin{enumerate}
    \item $\psi(b)$ and $\psi'(b)$ tend asymptotically towards zero;
    
    \item $g \equiv \delta + \psi(b)$ defines an asymptotically flat
    Riemannian metric on $\R^{3}$;
    
    \item $S \equiv \psi'(b)$ defines a symmetric tensor that is
    trace-free with respect to $g$;
    
    \item $g$ and $S$ satisfy the equations \eqref{eqn:simplextcon}. 
    \end{enumerate} \upshape

\smallskip \noindent The proof of this theorem will be presented in
the remaining sections of this article, and consists of essentially
two steps.  As outlined in the previous section,
\eqref{eqn:simplextcon} is not an elliptic system.  However, by
exploiting the elliptic properties of the equations, it is possible to
define a closely related system of equations, called the
\emph{associated system}, which is elliptic.  In it, the tensor $S$ is
decomposed into a sum of two components of the form $T+P(X)$, where
$T$ is a symmetric and trace-free tensor, $X$ is a 1-form and $P$ is
the adjoint of the divergence operator $S_{ab} \mapsto \nabla^{a}
S_{ab}$.  The system \eqref{eqn:simplextcon}, written in terms of this
decomposition, yields equations for $g$, $X$, and $T$ whose
linearization in the $g$ and $X$ directions is bijective (or near
enough to being bijective --- this will be cleared up in due course). 
Thus the Implicit Function Theorem can be invoked to find solutions
where the quantities $g$ and $X$ are expressed as functions of $T$,
which consists of the first step of the proof.  The second step is
then to show that all solutions of the associated system are also
solutions of the original system \eqref{eqn:simplextcon}.  The Author
wishes to thank H. Friedrich for suggesting this approach for solving
\eqref{eqn:simplextcon}.

The method outlined above for solving the extended constraints in the
time symmetric case is in fact a method for solving the usual vacuum
constraint equations in the time-symmetric case (namely the equation
$R(g)=0$, which follows from \eqref{eqn:simplextcon} by taking a
trace) because of the equivalence of the extended constraints and the
usual constraints described earlier.  The differences between this
method and the `classical' Lichnerowicz-York method for solving the
constraint equations are now readily apparent.  In the classical
method, one freely prescribes a metric $g_{0}$ on $\R^{3}$ and
considers the conformally rescaled metric $g = u^{4} g_{0}$, where $u
: \R^{3} \rightarrow \R$ is an unknown function.  One then reads the
equation $R(u^{4} g_{0})= 0$ as a semi-linear elliptic equation for
$u$.  In contrast, the present method treats the metric $g$ and the
one-form $X$ as the unknowns and leads to a quasi-linear elliptic
system for these quantities in terms of the freely prescribable
quantity $T$, which is a component of the curvature of the solution.

\medskip \noindent \scshape Remark: \upshape The Main Theorem does
\emph{not} fall into the domain of \emph{prescribed Ricci curvature}
as, for example, do the results of De Turck and his collaborators
\cite{deturck3,deturck2,deturck1,deturck5}.  In these papers, the
authors suppose a fixed symmetric tensor $S$ is given on a set
$\mathcal{O}$ and attempt to find conditions under which a metric $g$
exists on $\mathcal{O}$ so that $Ric(g) = S$.  In the Main Theorem, by
contrast, the tensor $S$ is itself an unknown quantity and only a
component is prescribed ahead of time by the free data.  Furthermore,
De Turck's results are local in nature since $\mathcal{O}$ is usually
an open set in $\R^{n}$, while the Main Theorem gives a global (though
perturbative) result.

\subsection{Formulating an Elliptic Problem}

The first task in the proof of the Main Theorem is to construct the
associated elliptic system that is to be solved by the Implicit
Function Theorem.  What is needed is a system of equations closely
related to \eqref{eqn:simplextcon} but that is elliptic.  To this end,
the the Ricci curvature operator in \eqref{eqn:simplextcon} will be
replaced by the reduced Ricci operator, which is elliptic as described
in Section 2.3.  Making this substitution is equivalent to assuming
\emph{a priori} that the harmonic coordinate condition is satisfied by
the metric $\delta + h$.  Of course, this assumption must be justified
later on; i.~e.\ it must be shown that $\delta + h$ does indeed
satisfy the harmonic coordinate condition, and this is the intent of
the second step of the proof of the Main Theorem.  The remaining
operator in \eqref{eqn:simplextcon} is underdetermined elliptic, and
an elliptic operator can be constructed from this by using a standard
technique known as the \emph{York decomposition} (see \cite{york} but
also \cite{cantor,com} for a thorough analysis of this method).  Write
a symmetric, trace-free tensor $S$ in terms of a 1-form $X$ and a
freely prescribed tensor symmetric $T$ as
$$S(h,X,T) = T^{\ast} + \mathcal{L}^{\delta + h}(X) \, .$$
where $T^{\ast} = T - \frac{1}{3} \mathrm{Tr}_{\delta+h}(T) (\delta +
h)$ is the trace-free part of $T$ and $\mathcal{L}^{\delta + h}(X)$ is
the conformal Killing operator with respect to the metric $\delta + h$
acting on $X$.  This is defined for a general metric $g$ by
$$\mathcal{L}^{g}_{ab}(X) = \nabla_{a} X_{b} + \nabla_{b} X_{a} -
\frac{2}{3} \nabla^{c} X_{c} g_{ab} \, ,$$
where $\nabla$ is the covariant derivative of the metric $g$.  The
reason for making this choice is that the composition of the
divergence operator in \eqref{eqn:simplextcon} and the conformal
Killing operator, that is the composite operator $\mathrm{div}_{g}
\circ \mathcal{L}^{g}$ given componentwise by
$$ X_{a} \mapsto \nabla^{a} ( \nabla_{a} X_{b} + \nabla_{b} X_{a} -
\frac{2}{3} \nabla^{c} X_{c} g_{ab}) = \nabla^{a} \nabla_{a} X_{b} +
\frac{1}{3} \nabla_{b} \nabla^{a} X_{a} +
R^{s}_{b}(g) X_{s} \, ,$$
is elliptic, as can easily be seen by computing its symbol or by
making the following observation.  It easy to compute that the
conformal Killing operator is the formal adjoint of the divergence
operator $S_{ab} \mapsto \nabla^{a} S_{ab}$ taking symmetric,
trace-free tensors to 1-forms.  Since this latter operator is
underdetermined elliptic, it is well-known that its adjoint is
overdetermined elliptic and that the composition of these two
operators as above is elliptic.

These considerations lead to the following definition of the
associated system, given here in index-free notation for ease of
presentation: 
\begin{equation}
    \begin{gathered}
	\Ric^{H} (\delta + h) = S(h,X,T) \\
	\div_{\delta + h} \circ S(h, X, T) = 0
    \end{gathered}
    \label{eqn:assoc}
\end{equation}
where $S(h,X,T)$ will be called the York operator.  As will be shown
in due course, the map defined by
\begin{equation}
    \Phi(h,X,T) \equiv \big(\Ric^{H}(\delta + h) - S( h,X,T) ,
    \div_{\delta + h} \circ S(h,X,T) \big)
    \label{eqn:map}
\end{equation}
on appropriate Banach spaces has a bounded, elliptic linearization in
the $h$ and $X$ directions and as a result, the Implicit Function
Theorem can be used to find solutions $h(T)$ and $X(T)$ as smooth
functions of sufficiently small tensors $T$.

\subsection{Choosing the Banach Spaces}

\label{sect:banach}

Before proceeding with the solution of the equations
\eqref{eqn:assoc}, it is necessary to specify in what Banach spaces of
tensors the equations are to be solved.  The notion of asymptotic
flatness in $\R^{3}$ should be encoded rigorously into the function
spaces by requiring that the relevant objects belong to a space of
tensors with built-in control at infinity.  Furthermore, the spaces
should be chosen to exploit the Fredholm properties of the operators
appearing in the map $\Phi$.  Both these ends will be served by
weighted Sobolev spaces, which are defined as follows.

Let $T$ be any tensor on $\R^{3}$.  (This tensor may be of any order
--- the norm $\Vert \cdot \Vert$ appearing in the following definition
is then simply the norm on such tensors that is induced from the
metric of $\R^{3}$.)  The $H^{k,\beta}$ Sobolev norm of $T$ is the
quantity
$$\Vert T \Vert_{H^{k,\beta}} = \left( \sum_{l=0}^{k} \int_{\R^{3}}
\Vert \nabla^{l} T \Vert^{2} \sigma^{-2(\beta-l) -3} \right)^{1/2} \,
,$$
where $\sigma(x) = ( 1+ r^{2} )^{1/2}$ is the \emph{weight function}
and $r^{2} = (x^{1})^{2} + (x^{2})^{2} + (x^{3})^{2}$ is the squared
distance to the origin.  Note that Bartnik's convention for describing
the weighted spaces is being used (the reason for this is
psychological: if $f \in H^{k,\beta}$ and $f$ is smooth enough to
invoke the Sobolev Embedding Theorem (see below), then $f(x) =
o(r^{\beta})$ as $r \rightarrow \infty$, which is easy to remember ---
see \cite{bartnik} for details).  An appropriate choice of $k$ and
$\beta$ for use in the Main Theorem will be made below.

The space of $H^{k,\beta}$ functions of $\R^{3}$ will be denoted by
$H^{k,\beta}(\R^{3})$ and the space of $H^{k,\beta}$ sections of a
tensor bundle $B$ over $\R^{3}$ will be denoted by $H^{k,\beta} (B)$. 
As an abbreviation, or where the context makes the bundle clear, such
a space may be indicated simply by $H^{k,\beta}$.  Note also that the
following convention for integration will be used in the rest of this
paper.  An integral of the form $\int_{\R^{3}} f$, as in the
definition above, denotes an integral of $f$ with respect to the
standard Euclidean volume form.  Integrals of quantities with respect
to the volume form of a different metric will be indicated explicitly,
as, for example, $\int_{\R^{3}} f \, \dif \vol$.

The spaces of $H^{k,\beta}$ tensors satisfy several important analytic
properties and the reader is asked to consult Bartnik's paper, or
others on the same topic \cite{bartnik,cantor,cbc,com}, for details. 
The three most important properties that will be used in the sequel
are the Sobolev Embedding Theorem, the Poincar\'e Inequality and
Rellich's Lemma; these will be restated here for easy reference.
\begin{enumerate}
    \itemsep = 1ex
    \item The Sobolev Embedding Theorem states that if $k >
    \frac{n}{2}$ and $T$ is a tensor in $H^{k,\beta}$, then $T$ is
    $C^{0}$.  Furthermore, if the weighted $C^{k}_{\beta}$ norm of a
    function $f$ is given by
    $$\Vert f \Vert_{C^{k}_{\beta}} = \sum_{l=0}^{k} \Vert \nabla^{l}
    f
    \sigma^{-\beta + l} \Vert_{0} \, ,$$
    where $\Vert f \Vert_{0} = \sup \{ \vert f(x) \vert \, : \, x \in
    \R^{3} \}$, then in fact, $T \in C^{0}_{\beta}$ and $ \Vert T
    \Vert_{C^{0}_{\beta}} \leq C \Vert T \Vert_{H^{k,\beta}}$,
    \item The Poincare Inequality states that if $\beta < 0$, then
    $$ \Vert f \Vert_{H^{0,\beta}} \leq C \Vert \nabla f \Vert_{H^{0,
    \beta
    -1}} \, ,$$
    whenever $f$ is a function in $H^{1,\beta}(\R^{3})$.  
    \item The Rellich Lemma states that the inclusion $H^{k,\beta}(B)
    \subseteq H^{k', \beta'}(B)$, for any tensor bundle $B$, is
    compact when $k' < k$ and $\beta' > \beta$.  In other words, if
    $T_{i}$ is a uniformly bounded sequence of tensors in
    $H^{k,\beta}$, then there is a subsequence $T_{i'}$ converging to
    a tensor $T$ in $H^{k', \beta'}$.
\end{enumerate} 

\noindent \scshape Remark: \upshape The constant $C$ appearing in the
estimates above is meant to depend only on the dimension $n$.  In the
remainder of this article, any constant depending only on $n$ will be
denoted by a generic $C$, unless it is important to emphasize
otherwise. \medskip 

In addition to the three properties above, two important results that
are valid in weighted Sobolev spaces will be needed in the sequel. 
The first concerns integration.

\medskip \noindent \bfseries Duality Lemma\mdseries: \itshape If $u
\in H^{l,\gamma}(\R^{3})$ and $v \in H^{l-2, -\gamma -3}$, then the
integral $\int_{\R^{3}} u \cdot v$ is well defined.  Furthermore, the
functional analytic dual space of $H^{0,\gamma}(\R^{3})$ is isomorphic
to $H^{0, -\gamma-3}(\R^{3})$ under the pairing $v \mapsto \phi_{v}$
where $\phi_{v}(u) = \int_{\R^{3}} u \cdot v$.\upshape

\medskip \noindent \itshape Proof: \upshape Choose $u$ and $v$ as in
the statement of the lemma.  Then by H\"older's inequality,
\begin{align*}
    \int_{\R^{3}} \vert u \cdot v \vert &\leq \int_{\R^{3}} \vert u
    \vert \sigma^{-\gamma - 3/2} \cdot \vert v \vert \sigma^{-
    (-\gamma-3) - 3/2} \\
    &\leq \left( \int_{\R^{3}} u^{2} \sigma^{-2\gamma - 3}
    \right)^{1/2} \left( \int_{\R^{3}} v^{2} \sigma^{-2(-\gamma-3) -
    3} \right)^{1/2} \\
    &< \infty \, .
\end{align*}
The product $u \cdot v$ is thus in $L^{1}$ and so its integral is well
defined.  The statement about duality follows from the Riesz
Representation Theorem for $L^{2}$ and the inequality above.  See
\cite{hebey,schwarz} for details.\hfill \QED

\medskip \noindent The second result concerns the Fredholm properties
of certain linear, elliptic partial differential operators on weighted
Sobolev spaces.

\medskip \noindent \bfseries Invertibility Theorem\mdseries: \itshape
Suppose $B$ is any tensor bundle over $\R^{3}$ and let $Q:
H^{k,\beta}(B) \rightarrow H^{k-2, \beta - 2}(B)$ be any linear,
second order, elliptic, homogeneous, partial differential operator
with constant coefficients mapping between weighted Sobolev spaces of
sections of $B$, and $k\geq 2$.  Then $Q$ is surjective if $\beta
\not\in \mathbf{Z}$ and $\beta > -1$ and injective if $\beta \not\in
\mathbf{Z}$ and $\beta < 0$.  It is thus bijective when $\beta \in
(-1,0)$.  The operator $Q$ is not Fredholm if $\beta \in \mathbf{Z}$. 
\upshape \medskip

\medskip \noindent \itshape Proof: \upshape The proof of this result
can be found in \cite{cbc}, but see also \cite{mcowen} for an
excellent discussion of the intuitive foundation underlying the theory
of elliptic operators on weighted spaces. \hfill \QED
   
\medskip \noindent \scshape Choice of Banach spaces \upshape \medskip

Denote by $S^{2}(\R^{3})$ the symmetric tensors over $\R^{3}$ and by
$\Lambda^{1}(\R^{3})$ the 1-forms of $\R^{3}$.  Solutions of the
associated system will be found in the following Banach spaces.  Pick
any $\beta \in (-1,0)$ and any $k \geq 4$; then
\begin{itemize}
    \item metrics $\delta + h$ will be found so that $h \in H^{k,\beta} 
    \big( S^{2}(\R^{3}) \big)$;
    
    \item 1-forms $X$ will be found in $H^{k-1, \beta-1} \big(
    \Lambda^{1}(\R^{3}) \big)$;
    
    \item tensors $T$ will be found in $H^{k-2, \beta-2} \big(
    S^{2}(\R^{3}) \big)$.
\end{itemize}

The preceding choice of Banach spaces will be justified in the next
section by showing that solutions of the associated system exist in
these spaces.  However, an argument can be made right now that
suggests that the spaces above are indeed the correct ones in which to
expect to find solutions.  First, in order to ensure that the metric
$\delta + h$ is asymptotically flat, $h$ must decay as $r \rightarrow
\infty$, and this holds by the Sobolev Embedding Theorem when $\beta
<0$.  Next, a non-trivial, asymptotically flat metric satisfying the
constraint equations must satisfy the Positive Mass Theorem \cite{sy}
and consequently must have non-zero ADM mass.  Thus the $r^{-1}$ term
in the asymptotic expansion of $h$ must be allowed to be non-zero,
which by the Sobolev Embedding Theorem imposes the further requirement
that $\beta > -1$.  Furthermore, $k \geq 4$ implies that the Sobolev
Embedding Theorem can be applied to the second derivatives of the
metric, and thus the curvature of the metric decays pointwise as $r
\rightarrow \infty$.  Finally, the $h$, $X$ and $T$ quantities are
chosen in different Sobolev spaces because of the differing numbers of
derivatives taken on these quantities in the associated system.  For
instance, the reduced Ricci curvature operator is homogeneous and of
degree two and thus sends a metric in $H^{k,\beta}$ to a tensor in
$H^{k-2, \beta-2}$.  The operator $S(h,X,T)$ is homogeneous but is
only of degree one in $X$ and of degree zero in $T$; it thus maps to
$H^{k-2, \beta -2}$ only when the weightings on $X$ and $T$ match
together properly and match the weighting on the metric $h$ as in the
choice above.

\subsection{First Attempt to Solve the Associated System}

\label{sect:solve}

The Implicit Function Theorem, the tool which will be used to solve
the associated system, is restated here for ease of reference.

\medskip \noindent \bfseries Implicit Function Theorem\mdseries:
\itshape Let $\Phi : A \times B \rightarrow C$ be a smooth map between
Banach spaces and suppose that $\Phi(0,0) = 0$.  If the restricted
linearized operator $D \Phi(0,0) \big|_{A \times \{0\}} : A
\rightarrow C$ is an isomorphism, then there exists an open set
$\mathcal{U} \subset B$ containing $0$ and a smooth function $\phi:
\mathcal{U} \rightarrow A$ with $\phi(0) = 0$ so that $\Phi \big(
\phi(b),b \big) = 0$.  \upshape \medskip

\noindent For an excellent discussion and proof of this theorem, see
\cite{amr}.  In order to use this theorem, let 
\begin{align*}
    A &= \big\{ (h,X) \in H^{k,\beta} \big( S^{2}(\R^{3}) \big) \times
    H^{k-1, \beta -1}(\Lambda^{1} \big( \R^{3}) \big) \big\}\\
    B &= \big\{ T \in H^{k-2,\beta-2} \big( S^{2}(\R^{3}) \big) \big\}
    \\
    C &= H^{k-2, \beta-2} \big( S^{2}(\R^{3}) \big) \times H^{k-3,
    \beta -3}(\Lambda^{1} \big( \R^{3}) \big) \, ;
\end{align*}
then the linearization of the operator $\Phi$ in the $A$ direction at the
origin must be calculated and its mapping properties understood.

The linearization of $\Phi$ is actually quite simple when evaluated at
the origin because the covariant derivative of the Euclidean metric is
trivial.  The only nonlinearities in $\Phi$ occur in the second order
terms of the reduced Ricci operator and in terms that are quadratic in
the derivatives of the metric (such as in products of Christoffel
symbols or in the connection terms).  It is thus easy to see that the
linearization of a covariant derivative operator at the Euclidean
metric is just the Euclidean derivative operator, and it is a
straightforward matter to deduce from the definition of the associated
system in \eqref{eqn:assoc} that the linearization of $\Phi$ in the $A
\times \{0\}$ direction is
\begin{equation}
	\mathrm{D} \Phi(0,0,0) \, (h,X,0) = \left( \begin{array}{c}
	-\frac{1}{2} \Delta h - \mathcal{L}(X) \\ 
	\mathrm{div} \circ  \mathcal{L}(X)  \end{array} \right) \, ,
    \label{eqn:lin}
\end{equation}
where $\Delta$ is the Euclidean Laplacian and $\mathcal{L}$ is the 
Euclidean conformal Killing operator.  

Denote by $P_{\delta}$ the operator $\mathrm{D} \Phi (0, 0, 0) (\cdot,
\cdot, 0)$.  It is a bounded linear operator between the appropriate
weighted Sobolev spaces because of the way in which the weights were
chosen in Section \ref{sect:banach}.  To determine whether
$P_{\delta}$ is an isomorphism, one appeals to the Invertibility
Theorem.  Recall that the weight $\beta$ in the domain spaces of
$P_{\delta}$ has been chosen between $-1$ and $0$.

\medskip \noindent \scshape Injectivity of $P_{\delta}$ \upshape 
\medskip

Suppose $(h, X)$ belong to the kernel of $P_{\delta}(h,X)$.  In other
words, $(h,X)$ solves the equation $P_{\delta}(h,X) = (0,0)$, or
\begin{gather*}
    -\frac{1}{2} \Delta h - \mathcal{L}(X) = 0 \\
    \mathrm{div} \circ \mathcal{L} (X) = 0 \, .
\end{gather*}
Since the operator $\mathrm{div} \circ \mathcal{L} : H^{k-1, \beta
-1}\big( \Lambda^{1}(\R^{3}) \big) \rightarrow H^{k-3, \beta -3} \big(
\Lambda^{1}(\R^{3}) \big)$ is a linear, elliptic, homogeneous,
constant coefficient operator of second order, the Invertibility
Theorem applies, and since $\beta -1 \in (-2,-1)$ when $\beta \in
(-1,0)$, it is thus injective.  Hence $X=0$.  The remaining equation
now reads $\Delta h = 0$ and again, since $\Delta : H^{k, \beta }
\big( S^{2}(\R^{3}) \big) \rightarrow H^{k-2, \beta -2} \big(
S^{2}(\R^{3}) \big)$ and $\beta \in (-1,0)$, $\Delta$ is an
isomorphism and thus $h = 0$.  Hence $P_{\delta}$ is injective.

\medskip \noindent \scshape Surjectivity of $P_{\delta}$ \upshape
\medskip

Although the operator $P_{\delta}$ is injective, it is \emph{not}
surjective.  First note that the Invertibility Theorem does not
\emph{guarantee} surjectivity in the same way that it guaranteed
injectivity.  To see this, attempt to solve the equations
$P_{\delta}(h,X) = (f,g)$ for any $f \in H^{k-2, \beta -2} \big(
S^{2}(\R^{3}) \big)$ and $g \in H^{k-3, \beta -3} \big(
\Lambda^{1}(\R^{3}) \big)$.  In other words, consider the system of
equations
\begin{gather*}
    -\frac{1}{2} \Delta h - \mathcal{L}(X) = f \\
    \mathrm{div}  \circ \mathcal{L}(X) = g \, .
\end{gather*}
Because $\beta - 1 \in (-2, -1)$, the operator $\mathrm{div} \circ
\mathcal{L}$ is not necessarily surjective according to the
Invertibility Theorem.  The full equations $P_{\delta}(h,X) = (f,g)$
can thus not necessarily be solved.

To show that $P_{\delta}$ actually does fail to be surjective, it is
necessary to show that the dimension of its cokernel in $H^{k, \beta}
\big( S^{2} (\R^{3}) \big) \times H^{k-1, \beta-3} \big( \Lambda^{1}
(\R^{3}) \big)$ is strictly greater that zero.  First, note that if
$X_{g}$ satisfies $\mathrm{div} \circ \mathcal{L}(X_{g}) = g$, then
the remaining equation $-\frac{1}{2} \Delta h = \mathcal{L}(X_{g}) +
f$ \emph{can} be solved by the Invertibility Theorem since the weight
$\beta$ is chosen such that $\Delta$ is an isomorphism.  Thus the
dimension of the cokernel of $P_{\delta}$ is equal to the dimension of
the cokernel of $\mathrm{div} \circ \mathcal{L}$ as an operator
between $ H^{k-1, \beta -1} \big( \Lambda^{1}(\R^{3}) \big)$ and $
H^{k-3, \beta -3} \big( \Lambda^{1}(\R^{3}) \big)$.

To characterize the cokernel of $\mathrm{div} \circ \mathcal{L}$, one
appeals to general, function-theoretic properties of linear, second
order, homogeneous, elliptic operators on weighted Sobolev spaces. 
The following lemma and its proof show how this is done.

\medskip \noindent \bfseries Cokernel Lemma\mdseries: \itshape Suppose
$B$ is any tensor bundle over $\R^{3}$ and let $Q : H^{k, \gamma}(B)
\rightarrow H^{k-2, \gamma -2}(B)$ be a linear, second order,
homogeneous, elliptic operator mapping between weighted Sobolev spaces
of sections of $B$ where $k \geq 2$ and $\gamma \not\in \mathbf{Z}$,
$\gamma < -1$.  The image of the operator $Q$ is the space:
\begin{equation}
    \mathrm{Im}(Q) = \left\{ w \in H^{k-2, \gamma -2}(B) \; : \;
    \int_{\R^{3}} \langle w, z \rangle = 0 \quad \forall \quad z \in
    \mathrm{Ker} \left( Q^{\ast}; -1-\gamma) \right) \right\} \, ,
    \label{eqn:coker}
\end{equation}
where the inner product $\langle \cdot , \cdot \rangle$ is induced on
$B$ from the Euclidean metric of $\R^{3}$, the operator $Q^{\ast}$ is
the formal adjoint of $Q$, and $\mathrm{Ker} (Q^{\ast}; -1-\gamma)$ is
its kernel as an operator from $H^{k, -1-\gamma}(B)$ to $H^{k-2,
-3-\gamma}(B)$.  \upshape

\medskip \noindent {\itshape Proof:} Denote the space on the right
hand side of equation \eqref{eqn:coker} by $W$.  Suppose that $k=2$
and consider first the containment $\mathrm{Im}(Q) \subseteq W$. 
Choose $Q(y) \in \mathrm{Im}(Q)$ and $z \in \mathit{Ker} (Q^{\ast};
-1-\gamma)$.  Since $Q(y) \in H^{2, \gamma-2}(B)$, the integral
$\int_{\R^{3}} \langle Q(y), z \rangle$ is well defined by the Duality
Lemma.  

\noindent \smallskip \scshape Claim: \upshape This integral equals
$\int_{\R^{3}} \langle y, Q^{\ast}(z) \rangle$. \smallskip

\noindent \itshape Proof: \upshape The equality of the integrals on
smooth, compactly supported sections of $B$ is true by definition of
the adjoint.  The equality of the integrals for $H^{k,\gamma}$
sections follows because $C^{\infty}_{c}$ sections of $B$ are dense in
$H^{k,\gamma}$ sections of $B$ \cite{bartnik}.  \smallskip

\noindent The integral $\int_{\R^{3}} \langle Q(y), z \rangle$ is thus
zero and so $Q(y) \in W$.

The reverse containment $W \subseteq \mathrm{Im}(Q)$ is proved as
follows.  Suppose $w_{0}$ belongs to $W$; thus, $w_{0} \in H^{0,
\gamma - 2}(B)$ and satisfies $\int_{R^{3}} \langle w_{0}, z \rangle =
0$ for all $z \in \mathit{Ker} (Q^{\ast}; -1-\gamma)$.  Suppose also
that $w_{0} \not\in \mathrm{Im}(Q)$.  Since $Q$ is elliptic,
$\mathrm{Im}(Q)$ is closed; thus by the Hahn-Banach theorem, there
exists a linear functional $\phi$ on $H^{0, \gamma - 2}(B)$ so that
$\phi(w_{0}) \neq 0$ but $\phi \big|_{\mathrm{Im} (Q)} = 0$.  Again by
the Duality Lemma, there is a unique $z_{0} \in H^{0, -1-\gamma}(B)$
so that $\phi(w) = \int_{\R^{3}} \langle w, z_{0} \rangle$ for all $w
\in H^{0, \gamma-2}(B)$.  Therefore, $\phi \big|_{\mathrm{Im}(Q)} = 0$
implies that
\begin{align*}
    0 &= \phi(Q(y)) \\
    &= \int_{\R^{3}} \langle z_{0}, Q(y) \rangle\\
    &= \int_{\R^{3}} \langle Q^{\ast}(z_{0}), y \rangle
\end{align*}
for all $y \in H^{2, \gamma}(B)$.  Thus $Q^{\ast} (z_{0}) = 0$ or
$z_{0} \in \mathrm{Ker}(Q^{\ast}; -1-\gamma)$.  But now, the
assumptions $\phi(w_{0}) \neq 0$ and $\int_{\R^{3}} \langle w_{0}, z
\rangle = 0$ for all $z \in \mathrm{Ker} (Q^{\ast}; -1-\gamma)$ are
mutually contradictory.  Thus it must be that $w_{0} \in
\mathrm{Im}(Q)$.  Finally, the extension to $k >2$ follows in a
similar manner by standard functional analysis.  \hfill \QED \medskip

Apply this theorem to the operator $Q = \mathrm{div} \circ
\mathcal{L}$ with $\gamma = \beta -1$.  Now, $Q^{\ast} = Q$, so in
order to solve the equation $\mathrm{div} \circ \mathcal{L} (X) = g$,
the tensors $g$ must satisfy the constraints
$$ \int_{\R^{3}} g_{a}Y^{a} = 0 \, ,$$
where $Y$ is any tensor in the kernel of the operator $\mathrm{div}
\circ \mathcal{L}$ in the space $H^{k-1,-1-\gamma} \big( \Lambda^{1}
(\R^{3}) \big)$.

The kernel of $\mathrm{div} \circ \mathcal{L}$ is well known and
consists of 1-forms dual to the the conformal Killing fields of
$\R^{3}$.  There are precisely ten linearly independent families of
such vector fields: the translation vector fields, the rotation vector
fields, the dilation field and three so-called \emph{special}
conformal Killing fields (these correspond to transformations of the
form $i \circ T \circ i$, where $i$ is the inversion with respect to
the unit circle and $T$ is a translation).  The asymptotic behaviour
of these vector fields can thus be computed exactly: the translations
have constant norm, the rotations and dilations have norm growing
linearly in the distance from the origin, and the special vector
fields have quadratic growth in the distance from the origin.  Since
$-1-\gamma \in (0,1)$ when $\beta \in (-1,0)$, the only 1-forms dual
to the conformal Killing fields in $H^{k-1,-1-\gamma} \big(
\Lambda^{1} (\R^{3}) \big)$ are thus those spanned by the translation
1-forms $\dif x^{1}$, $\dif x^{2}$ and $\dif x^{3}$.  Consequently,
the image of $Q = \mathrm{div} \circ \mathcal{L}$ in the space
$H^{k-3, \gamma -2} \big( \Lambda^{1} (\R^{3}) \big)$ can be
characterized as follows:
$$ \mathrm{Im}(\mathrm{div} \circ \mathcal{L}) = \left\{ g \in H^{k-3,
\beta-3} \big( \Lambda^{1} (\R^{3}) \big) \: : \: \int_{\R^{3}} g_{a}
= 0 \, , \, \, a=1,2,3 \right\}
\, ,$$
where $g_{a}$ are the components of $g$ in the standard coordinates of
$\R^{3}$.

The conclusion that can be drawn from the analysis of this section is
that the equation $\Phi(h, X, T) = (0,0)$ is \emph{not} solvable near
$(0, 0, 0)$ using the Implicit Function Theorem.  The non-surjectivity
of the linearized operator at $(0,0,0)$ is the essential obstruction. 
The best that can be achieved using the Implicit Function Theorem is
thus that the equation $\Phi(h, X, T) = (0,0)$ can be solved \emph{up
to} a term that is transverse to the space $\mathit{Im}(\div \circ
\mathcal{L})$.  It will turn out that this is nevertheless sufficient
for solving the full equations as a result of the compatibility
conditions built into the equations.  But in order to show this, the
associated system defined in the previous section must be modified
somewhat.

\subsection{Reestablishing Surjectivity and Solving the Associated
System}

\label{sect:surj}

In order to modify the associated system appropriately, first note
that $H^{k-3, \beta -3} \big( \Lambda^{1} (\R^{3}) \big)$ can be
written as $\mathit{Im}(\div \circ \mathcal{L}) \oplus W$ in many
different ways; but in each case, $W$ is a three dimensional subspace
of $H^{k-3, \beta -3} \big( \Lambda^{1} (\R^{3}) \big)$ whose members
do not integrate to zero upon taking the Euclidean inner product with
the translation 1-forms.  One such choice is
$$W = \mathrm{span} \left\{ \phi \, \dif x^{a} \right\} _{a = 1,2,3}
\, ,$$
where $\phi$ is any smooth, positive function of compact support whose
integral over $\R^{3}$ is equal to 1.

Again, denote the domain space of the operator $\Phi$ by $A$.  The
previous paragraph suggests that one should attempt to construct a new
associated operator $\Phi'$ that extends $\Phi$ in such a way that
$\Phi' : A \times \R^{3} \rightarrow \mathrm{Im}(P_{\delta}) \oplus
W$, where the additional $\R^{3}$ factor in the domain should map
under the linearization $\mathrm{D} \Phi'$ at the solution $(0, 0, 0;
0) \in A \times \R^{3}$ onto the $W$ factor in the image.  If such a
construction is possible, then the equation $\Phi'(h,X,T; \lambda) =
(0,0)$ can be solved using the Implicit Function Theorem.

Construct the operator $\Phi' : A \times \R^{3} \rightarrow H^{k-3,
\beta -3} \big( \Lambda^{1} (\R^{3}) \big)$ according to the
prescription
\begin{equation}
    \Phi'(h,X,T;\lambda) = \left( Ric^{H}(\delta + h) - S(h,X,T),
    \, \mathrm{div}_{\delta + h} \circ S(h,X,T) - \sum_{a=1}^{3}
    \lambda_{a} \phi \, \dif x^{a} \right) \, ,
    \label{eqn:prime}
\end{equation}
where, as before, $\mathrm{Ric}^{H}$ is the reduced Ricci operator and
$S(\cdot , \cdot , \cdot)$ is the York operator.  The linearization of
$\Phi'$ at $(0, 0,0;0)$ in the directions transverse to the $T$
direction is easily seen to be
\begin{equation}
    \mathrm{D} \Phi'(\delta, 0,0;0)(h,X,0;\lambda) = \left(
    -\frac{1}{2} \Delta h - \mathcal{L}(X), \, \mathrm{div} \circ
    \mathcal{L}(X) - \sum_{a=1}^{3} \lambda_{a} \phi \, \dif x^{a}
    \right) \, .
    \label{eqn:linprime}
\end{equation}
Denote this new operator by $P_{\delta}'$.  It is still bounded
because $\phi$ has compact support, and it is now also bijective by
the following arguments.

\medskip \noindent \scshape Injectivity of $P_{\delta}'$ \upshape 
\medskip 

Suppose $P_{\delta}'(h,X;\lambda) = (0,0)$.  Integrate the components
of the second equation; by the divergence theorem for the Euclidean
metric (valid because constant functions can be integrated against
$H^{k-3, \beta-3}$ functions when $\beta \in (-1,0)$ according to the
Duality Lemma), the divergence terms integrate to zero, yielding
$\lambda_{a}=0$ for all $a$.  The argument that both $X$ and $h$ are
then equal to zero follows as in Section \ref{sect:solve}.

\medskip \noindent \scshape Surjectivity of $P_{\delta}'$ \upshape 
\medskip

Suppose that $P_{\delta}'(h,X; \lambda) = (f,g)$.  First choose the
components $\lambda_{a}$ so that
$$\int_{\R^{3}} \big( g_{a} + \lambda_{a} \phi \big) = 0
$$
for each $a$.  The equation $\mathrm{div} \circ \mathcal{L}(X) = g -
\sum_{a=1}^{3} \lambda_{a} \phi \, \dif x^{a}$ can then be solved for
$X_{g}$ according to the characterization of the image of the operator
$\mathrm{div} \circ \mathcal{L}$ from the previous section.  The
remaining equation $-\frac{1}{2} \Delta h = \mathcal{L}(X_{g}) + f$
can then be solved because $\beta \in (-1,0)$ makes $\Delta$ an
isomorphism.

\medskip

The Implicit Function Theorem can now be invoked to solve the equation
$\Phi'(h,X,T;\lambda) = (0,0)$ near $(0, 0,0;0)$.  To be precise,
there is a neighbourhood $\mathcal{U} \subset H^{k-2, \beta -2} \big(
S^{2}(\R^{3}) \big)$ with the following property.  If $T \in
\mathcal{U}$, then there is a metric $\delta + h(T)$ with $h(T) \in
H^{k,\beta}\big( S^{2}(\R^{3}) \big)$, a covector field $X(T) \in
H^{k-1,\beta-1}\big( \Lambda^{1}(\R^{3}) \big)$, and three real
numbers $\lambda_{a}(T)$ so that $\Phi' \big( h(T), X(T), T;
\lambda(T) \big) = (0,0)$.  Furthermore, the various functions $T
\mapsto h(T)$, etc.\, are smooth in the appropriate Banach space
norms.  In particular, there exists a constant $C$ so that
\begin{equation}
    \begin{gathered}
	\Vert h \Vert_{H^{k, \beta}} \leq C \Vert T \Vert_{H^{k-2,
	\beta - 2}} \\
	\Vert X \Vert_{H^{k-1, \beta -1}} \leq C \Vert T
	\Vert_{H^{k-2, \beta -2}} \\
	\Vert \lambda \Vert_{\R^{3}} \leq C \Vert T \Vert_{H^{k-2,
	\beta - 2}} \, ,
    \end{gathered}
    \label{eqn:contest}
\end{equation}
where $\Vert \cdot \Vert_{\R^{3}}$ denotes the standard Euclidean norm
of $\R^{3}$, as long as $T \in \mathcal{U}$.

\subsection{Satisfying the Harmonic Coordinate Condition}

Section \ref{sect:surj} shows how the associated system
\eqref{eqn:assoc} can be modified in such a way that it can be
solved using the Implicit Function Theorem.  This procedure results in
a family of solutions of the equations
\begin{equation}
    \begin{gathered}
	\mathit{Ric}^{H}(\delta + h) = S(h,X,T) \\
	\div_{\delta + h} \circ S(h,X,T) = \lambda \phi \, ,
    \end{gathered}
    \label{eqn:tempeqn}
\end{equation}
where $\lambda = \sum_{a=1}^{3} \lambda_{a} \, \dif x^{a}$.  It
remains to show whether the original equations \eqref{eqn:simplextcon}
are satisfied by the solution $\delta + h$ and $S(h,X,T)$.  This will
be done by showing that the compatibility conditions built into the
extended constraint equations (i.~e.\ the Bianchi identity only, since
the time-symmetric assumption has eliminated the other compatibility
condition) actually ensure that if $(h, X, T;\lambda)$ solves
\eqref{eqn:tempeqn}, then $\lambda = 0$ and $h + \delta$ satisfies the
harmonic coordinate condition.  Therefore $Ric^{h}(\delta + h) =
Ric(\delta + h)$ and solutions of \eqref{eqn:tempeqn} are indeed
solutions of the full equations.

To prove this claim, assume instead that both $\lambda$ and the
quantities $\Gamma^{a}$ are nonzero.  Argue towards a contradiction as
follows.  First, write $g = \delta + h$ for short.  The Bianchi
identity $\div_{g} \big( \Ric(g) - \frac{1}{2} R(g) g \big) = 0$,
applied to equation \eqref{eqn:redric} defining the reduced Ricci
operator yields the identity
\begin{equation*}
    0 = \big( R^{H}_{ab} - \frac{1}{2} R^{H} g_{ab}
    \big)_{;}^{\rule{0.5ex}{0ex}a} = \big( \Gamma_{a;b} + \Gamma_{b;a}
    - \Gamma^{c}_{ \rule{0.5ex}{0ex};c} h_{ab}
    \big)_{;}^{\rule{0.5ex}{0ex}a}
\end{equation*}
which is equivalent to
\begin{equation}
    \Gamma_{b;a}^{\rule{2.5ex}{0ex} a} + R_{b}^{a} \Gamma_{a} = 2 \phi
    \lambda_{a}\, ,
    \label{eqn:neweqn}
\end{equation}
after using the modified associated system and commuting covariant
derivatives appropriately.  If $Q_{h}$ denotes the operator $u_{a}
\mapsto \Delta_{\delta + h} u_{a} + [Ric(\delta + h)]_{a}^{b} u_{b}$,
then \eqref{eqn:neweqn} asserts that $2 \phi \lambda_{a}$ is in the
image of $H^{k-1,\beta-1} (\Lambda^{1} (\R^{3}))$ under $Q_{h}$,
because $h \in H^{k, \beta} \big( S^{2} ( \R^{3}) \big)$ and the
$\Gamma^{a}$ are obtained from $\delta + h$ by differentiation.  This,
however, can be shown to violate the following basic result about
elliptic operators.

\medskip \noindent \bfseries Stability Lemma\mdseries: \itshape Let
$B$ be a tensor bundle over $\R^{3}$ and let $Q_{\eps}:H^{l,\gamma}
(B) \rightarrow H^{l-2,\gamma-2} (B)$, $\eps \in [0,1]$, be a
continuous family of linear, homogenous, second order, elliptic
operators, for all $\gamma < -1$.  Furthermore, suppose $Q_{\eps}$ is
uniformly injective for any $\eps$ whenever $\gamma < -1$; i.~e.\ for
each $\gamma \not\in \mathbf{Z}$, $\gamma < -1$, there is a constant
$C$ independent of $\eps$ so that $\Vert Q_{\eps}(y)
\Vert_{H^{l-2,\gamma-2}} \geq C \Vert y \Vert_{H^{k, \gamma}}$.  If $z
\not\in \mathrm{Im}(Q_{0})$, then there exists $\eps_{0} > 0$ so that
$z \not\in \mathrm{Im}(Q_{\eps})$ for all $\eps < \eps_{0}$.  \upshape

\medskip \noindent \itshape Proof: \upshape Suppose the contrary; then
for some $\gamma < -1$, there exists a sequence $\eps_{i} \rightarrow 0$
and a sequence $y_{i} \in H^{l,\gamma}(B)$ so that $z =
Q_{\eps_{i}}(y_{i})$.  By the uniform injectivity of $Q_{\eps}$, $\Vert
y_{i} \Vert_{H^{l,\gamma}} \leq C \Vert z \Vert_{H^{l-2, \gamma-2}}$
and is thus uniformly bounded.  By Rellich's Lemma, there exists a
subsequence $y_{i'}$ which converges to an element $y$ in $H^{l-1,
\gamma + \rho}$, where $\rho$ is small enough so that $\gamma + \rho <
-1$.  Again, by uniform injectivity,
\begin{align*}
    \Vert y_{i'} - y_{j'} \Vert_{H^{l, \gamma + \rho}} &\leq C \Vert
    Q_{\eps_{i'}}(y_{i'} - y_{j'}) \Vert_{H^{l-2, \gamma + \rho -2}}
    \\
    &\leq C \Vert (Q_{i'} - Q_{j'})(y_{j'}) \Vert_{H^{l-2, \gamma + 
    \rho -2}} \\
    &\leq C \Vert Q_{i'} - Q_{j'} \Vert_{op} \cdot \Vert y_{j'}
    \Vert_{H^{l, \gamma+\rho}} \\
    &\leq C \Vert Q_{i'} - Q_{j'} \Vert_{op} \cdot \Vert y_{j'}
    \Vert_{H^{l, \gamma}} \\
    &\longrightarrow 0 \, ,
\end{align*}
by the continuity of $Q_{\eps}$ and the uniform boundedness of $y_{i}$. 
Here, $\Vert \cdot \Vert_{op}$ denotes the relevant operator norm.  
The subsequence $y_{i'}$ is thus Cauchy in the $H^{l,\gamma+\rho}$ 
norm and so $y_{i'} \rightarrow y$ in this norm.  But now,
$$z = \lim_{i' \rightarrow \infty} Q_{\eps_{i'}} (y_{i'}) = Q_{0}(y)
\, ,$$
contradicting the fact that $z \not\in \mathrm{Im}(Q_{0})$.
\hfill \QED

\medskip

In order to derive a contradiction from \eqref{eqn:neweqn} using this
lemma, the uniform injectivity of $Q_{h}$ must be established and it
must be shown that $\phi \lambda_{a}$ does not belong to the image of
$Q_{0}$.

\medskip \noindent \scshape Uniform Injectivity of $Q_{h}$ \upshape
\medskip

Suppose that $Q_{h}(u) = 0$ for $u \in H^{k-1, \gamma} \big(
\Lambda^{1} (\R^{3}) \big)$ where $\gamma < -1$.  In other words,
$\Gamma_{b;a}^{\rule{2.5ex}{0ex} a} + R_{b}^{a} \Gamma_{a} = 0$.  From
this, one easily deduces
\begin{equation}
    - \Delta_{g} \Vert u \Vert^{2} = 2 \big( R_{ab} u^{a} u^{b}
    - \Vert \nabla u \Vert^{2} \big) \, .
    \label{eqn:contrad}
\end{equation}

Before continuing, recall the following facts about Green's identity
in weighted Sobolev spaces.  If functions $u$ and $v$ are chosen such
that $v \in H^{k, \gamma}(\R^{3})$ and $u \in H^{k, -1 - \gamma}
(\R^{3})$ for some $\gamma$, then the integrals appearing Green's
identity for a general metric $g$ on a large ball $B_{r}$, that is
\begin{equation}
    \int_{B_{r}} u \Delta_{g} v \, \dif \mathrm{Vol}_{g} +
    \int_{B_{r}} \nabla u \cdot \nabla v \, \dif \mathrm{Vol}_{g} =
    \int_{\partial B_{r}} u \frac{\partial v}{\partial n} \, \dif
    \mathrm{A}_{g} \, ,
    \label{eqn:approxgreen}
\end{equation}
where $\dif \mathrm{A}_{g}$ is the area form of the metric $g$, are
all well defined as $r \rightarrow \infty$.  Thus by applying a
density argument as in the proof of the Cokernel Lemma, one
can conclude that
$$\int_{\R^{3}} u \Delta_{g} v \, \dif \mathrm{Vol}_{g} +
\int_{\R^{3}} \nabla u \cdot \nabla v \, \dif \mathrm{Vol}_{g} = 0 \,
, $$
in the limit of \eqref{eqn:approxgreen} as $r \rightarrow \infty$.

With this in mind, integrate both sides of equation
\eqref{eqn:contrad} against the volume form of the metric $g = \delta
+ h$ to obtain
\begin{equation}
    -\frac{1}{2} \int_{\R^{3}} \Delta_{g} \Vert u \Vert^{2} \, \dif
    \mathrm{Vol}_{g} = \int_{\R^{3}} R_{ab} u^{a} u^{b} \, \dif
    \mathrm{Vol}_{g} - \int_{\R^{3}} \Vert \nabla u \Vert^{2} \, \dif
    \mathrm{Vol}_{g} \, .
    \label{eqn:green}
\end{equation}
Since $u \in H^{k-1, \gamma}$, Green's Identity can be applied to the
left hand side of \eqref{eqn:green} when $1 \in H^{k-1, -\gamma - 1}$. 
This is true since $\gamma < -1$; thus the integral of the left hand
side of \eqref{eqn:green} is zero.  Consequently,
\begin{align}
    0 &\leq \int_{\R^{3}} \Vert \Ric (g) \Vert \, \Vert u \Vert^{2}
    \, \dif \vol - \int_{\R^{3}} \Vert \nabla u \Vert^{2} \, \dif \vol
    \notag \\
    &\leq \int_{\R^{3}} \Vert \Ric (g) \Vert \, \Vert u \Vert^{2} \,
    \dif \vol - C \int_{\R^{3}} \Vert \, \nabla \Vert u \Vert \,
    \Vert^{2} \, \dif \vol
    \label{eqn:ineq}
\end{align}
for some constant $C$, by the Cauchy-Schwartz inequality and
straightforward algebra.  Next, assume that $h$ is small in a
pointwise sense (this assumption follows from the Sobolev Embedding
Theorem if $h$ is sufficiently small in the $H^{k,\beta}$ norm and $k
> \frac{3}{2}$).  In fact, assume that $h$ is sufficiently close to
$0$ so that all norms, derivatives and volume forms of the metric $g$
can be replaced by their Euclidean counterparts (at the expense of
changing $C$ of course).  Finally, since $\Vert u \Vert$ is a scalar
function, the derivative operator in \eqref{eqn:ineq} can be replaced
by the Euclidean derivative operator without introducing lower order
terms.  Thus, there exists a new constant $C$ so that the estimate
\begin{equation}
    0 \leq \int_{\R^{3}} \Vert \Ric (g) \Vert \, \Vert u \Vert^{2} -
    C \int_{\R^{3}} \Vert \, \nabla \Vert u \Vert \, \Vert^{2}
    \label{eqn:ineq2}
\end{equation}
holds, where the norms and derivatives appearing here are those of the
Euclidean metric.  Next, $\mathrm{Ric}(g) \in H^{k-2, \beta-2}$
because $g - \delta \in H^{k,\beta}$.  But since $k > \frac{7}{2}$,
the Sobolev Embedding Theorem gives $\mathrm{Ric}(g) \in C^{0}_{-
\beta + 2}$.  That is,
$$\sup_{\R^{3}} \left \Vert \mathrm{Ric}(g) \cdot \sigma^{- \beta + 2}
\right \Vert \leq C < \infty \, ,$$
which implies that
$$\sup_{\R^{3}} \left \Vert \mathrm{Ric}(g) \cdot \sigma^{2}
\right \Vert \leq C < \infty \, ,$$
since $\beta < 0$.  Finally, apply the Poincar\'e inequality for
weighted Sobolev norms to the function $\Vert u \Vert$ to deduce
\begin{align}
    \int_{\R^{3}} \Vert \mathrm{Ric}(g) \Vert \, \Vert u \Vert^{2}
    &\leq \Vert \mathrm{Ric}(g) \cdot \sigma^{2} \Vert_{\, 0}
    \int_{\R^{3}} \Vert u \Vert^{2} \sigma^{-2} \notag \\
    &\leq C \Vert \mathrm{Ric}(g) \Vert_{\, C^{0}_{-2}} \int_{\R^{3}}
    \Vert \, \nabla \Vert u \Vert \, \Vert^{2} \notag \\
    &\leq C \Vert g - \delta \Vert_{C^{2}_{0}} \int_{\R^{3}} \Vert \,
    \nabla \Vert u \Vert \, \Vert^{2} \notag \\
    &\leq C \Vert h \Vert_{H^{k,\beta}} \int_{\R^{3}}\Vert \, \nabla
    \Vert u \Vert \, \Vert^{2}
    \label{eqn:poincare}
\end{align}
again by the Sobolev Embedding Theorem and the fact that $\beta <0$. 
Using \eqref{eqn:poincare} in inequality \eqref{eqn:ineq2} leads to
the contradiction because the preceding estimates imply
\begin{equation*}
    0 \leq \left( C \Vert h \Vert_{H^{k,\beta}} - 1 \right)
    \int_{\R^{3}} \Vert \, \nabla \Vert u \Vert \, \Vert^{2} \, ,
\end{equation*}
while if $\Vert h \Vert_{H^{k,\beta}}$ is sufficiently small, the
right hand side above is clearly negative.  Avoiding this
contradiction requires $\nabla \Vert u \Vert = 0$.  But since the
Sobolev Embedding Theorem applied to $u \in H^{k-1,\gamma}$ shows that
$\Vert u \Vert$ decays at infinity when $\gamma <-1$, it must be true
that $u = 0$.

The operator $Q_{h}$ acting on $H^{k-1,\gamma}$ 1-forms is injective
for all $\gamma < -1$ whenever $h$ is sufficiently close to zero in
the $H^{k,\beta}$ norm.  The uniform injectivity follows in the
standard way from the injectivity of each $Q_{h}$ and the fact that
the constant in the elliptic estimate for these operators is
independent of $h$, again provided $h$ is sufficiently near to $0$.

\medskip \noindent \scshape Image of $Q_{0}$ \upshape \medskip

The $\phi \lambda$ term in \eqref{eqn:tempeqn} was specifically chosen
in Section 3.5 to satisfy the integral condition $\int_{\R^{3}}
\langle \lambda \phi, \dif x^{b} \rangle \neq 0$ (since $\lambda_{a} =
0$ for all $a$).  This condition ensures that indeed $2 \phi
\lambda_{a}$ is not in the image of the operator $Q_{0} =
\Delta_{\delta}$ acting on the space of $H^{l,\gamma}$ 1-forms of
$\R^{3}$ because the image of $\Delta_{\delta}$ in $H^{l,\gamma}$ for
$\gamma < -1$ is perpendicular to the harmonic polynomials of degree
less than the nearest integer less than $\gamma$, and this always
includes the constants.

\medskip 

The Stability Lemma thus applies to equation \eqref{eqn:neweqn} and
implies that $\phi \lambda$ can not be in the image of $Q_{h}$ when
$h$ is sufficiently small in the $H^{k,\beta}$ norm, unless of course
$\lambda = 0$.  Now, by the injectivity of the operator $Q_{h}$, this
in turn implies that $\Vert \Gamma \Vert = 0$, or that $\Gamma^{a} =
0$ for each $a$.  Consequently, the harmonic coordinate condition for
the metric $\delta + h$ is satisfied, and as indicated earlier, this
implies that the the metric $\delta + h$ and the tensor $S(h,X,T)$
satisfy the time-symmetric extended constraint equations.  This
completes the proof of the Main Theorem.  \hfill \QED

\medskip \noindent \bfseries Acknowledgements\mdseries: I would like
to thank Helmut Friedrich for suggesting the topic of the conformal
constraint equations to me as well as for providing me with guidance
and insight.  I would also like to thank Piotr Chru\'sciel, J\"org
Frauendiener and Niall O'Murchadha for their valuable comments and
suggestions.

\newpage

\bibliography{grav}
\bibliographystyle{amsplain}

\end{document}